\documentclass[epj,nopacs]{svjour}

\usepackage{graphicx}% Include figure files
\usepackage{bm}% bold math
\usepackage{graphicx}
\usepackage{amssymb}
\usepackage{amsmath}
\usepackage{color}
\usepackage{mathrsfs}
\usepackage[english]{babel} 

\newcommand{\bs}[1]{\bf{#1}}                      % Vectors: bold
\usepackage{dcolumn}

\definecolor{orange}{rgb}{1,0.5,0}

\newcommand{\mbs}[1]{{\boldsymbol{#1}}}

\begin{document}

\title{Focusing and splitting of particle streams in microflows via viscosity gradients}% Force line breaks 

\titlerunning{ Focusing and splitting of particle streams via viscosity gradients }
\authorrunning{ Matthias Laumann and Walter Zimmermann}

\author{Matthias Laumann and Walter Zimmermann}
\institute{Theoretische Physik I, Universit\"at Bayreuth, 95440 Bayreuth, Germany; E-mail: walter.zimmermann@uni-bayreuth.de}

\date{Received: date / Revised version: date}

\abstract{
Microflows are intensively used for investigating and controlling the dynamics of particles, including soft particles  such as biological cells  and capsules. A classic result is the tank-treading motion of elliptically deformed soft particles in linear shear flows, which do not migrate across straight stream lines in the bulk.  However, soft particles migrate across straight streamlines in Poiseuille flows. In this work we describe a new mechanism of cross-streamline migration of soft particles. If the viscosity varies perpendicular to the stream lines then particles migrate across stream lines towards regions  of a lower viscosity, even in linear shear flows. An interplay with the repulsive particle-boundary interaction causes then focusing of particles in linear shear flows with the attractor stream line closer to the wall in the low viscosity region. Viscosity variations perpendicular to the stream lines  in Poiseuille flows leads either to a shift of the particle attractor or even to a  splitting of particle attractors, which may give rise to interesting applications for particle separation. The location of attracting streamlines  depend on the particle properties, like their size and elasticity. The cross-stream migration induced by  viscosity variations is explained by analytical considerations, Stokesian dynamics simulations with a generalized Oseen tensor and Lattice-Boltzmann simulations.
}

\maketitle

%============================================================================================
\section{Introduction}
%============================================================================================

\begin{figure}[t!]
\begin{center}
\includegraphics[width=0.98\columnwidth]{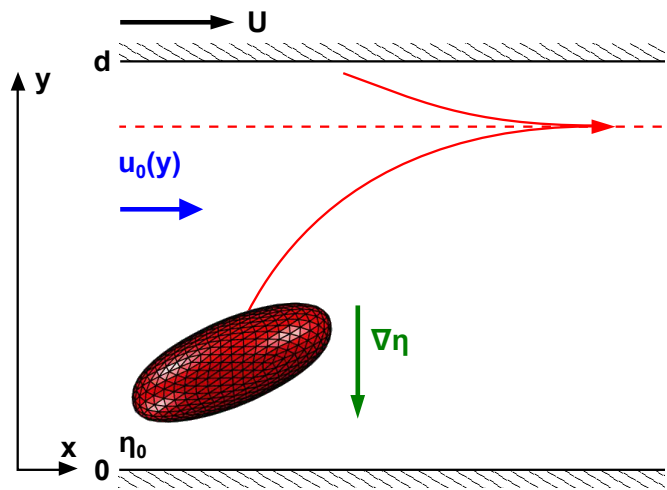}
\end{center}
\caption{The two solid lines sketch two trajectories of a soft capsule (enlarged) 
in a shear flow  ${\bf u}_0(y)$ driven by a moving upper boundary.
The shear viscosity  of the fluid increases from top to bottom (for instance induced by 
a temperature gradient) and the capsule migrates towards the region  of low 
viscosity. Along the attractor (dashed line) the migration to a smaller viscosity  is in
 balance with the particle repulsion by the upper boundary.
}
\label{fig_Sketch_tra}
\end{figure}

The success of the interdisciplinary field of microfluidics and its numerous applications in life science 
and applications are  based also on a
thorough  understanding of the  dynamics of particles and their distribution in microflows. 
\cite{Quake:2005.1,Popel:2005.1,Karimi:2013.1,Kumar_S:2015.1,DiCarlo:2014.1,Secomb:2017.1} 
One of the important applications is particle sorting where besides structured channels 
also optical, electrical or magnetic fields are used.\cite{Karimi:2013.1} 
Several sorting strategies rely merely on the interplay between basic hydrodynamics 
of microflows and  particle properties, 
that cause, for instance,  cross-streamline migration (CSM) of particles. 
CSM may depend  on fluid inertia,\cite{DiCarlo:2014.1} on particle deformability,
\cite{Misbah:1999.1,Seifert:1999.1,Viallat:2002.1,Podgorski:2013.1,Leal:1980.1,Chakraborty:2015.1,BarthesBiesel:1980.1,Kaoui:2008.1,Misbah:2008.1,Bagchi:2008.1,Laumann:2017.2}  on channel modulations\cite{Foerster:2018.1} 
or on non-Newtonian fluid effects.\cite{DAVINO:2010.1,Dan:2018.1,Giudice:2017.1,Lu:2017.1,Faridi:2017.1,Avino:2017.1}
In non-Newtonian flows the action of elastic effects and spatially varying shear viscosity on particles
come often simultaneously into play, but little is known about the action of a spatially dependent  shear viscosity
on the particle dynamics alone.
We describe in this work a surprising  viscosity-gradient driven CSM and the resulting focusing 
of soft-particles, which occurs even in linear shear flows as indicated in Fig.~\ref{fig_Sketch_tra}.

Segr\'e and Silberberg found quite early
that rigid particles can migrate across straight streamlines 
to off-center streamline positions in pipe flows \cite{Silberberg:1961.Nat}.
This type of cross-streamline migration (CSM) is inertia driven in the range 
of intermediate Reynolds number ($\sim 1 < Re < \sim 100$) and it
is extensively used for sorting of rigid particles (see e.g. Ref. \cite{DiCarlo:2014.1}). 
In contrast, deformable particles like capsules or cells show
CSM already on the scale of microchannels and in the limit of Stokes flows at very small values of the Reynolds number.
The tank-treading motion of vesicles or  capsules causes near walls the so-called lift force
that drives them away from channel walls in Poiseuille and linear shear 
flows  \cite{Misbah:1999.1,Seifert:1999.1,Viallat:2002.1,Podgorski:2013.1}.
 Further away from the walls in  Poiseuille flows one still has a spatially varying shear rate, which breaks  the fore-aft symmetry
of  deformed particles, so that dumbbells \cite{Armstrong:1982.1,Brunn:1983.1,Freed:1985.1},
droplets \cite{Leal:1980.1,Chakraborty:2015.1}, vesicles and capsules \cite{BarthesBiesel:1980.1,Kaoui:2008.1,Misbah:2008.1,Bagchi:2008.1} 
exhibit bulk CSM, even in unbounded Poiseuille flows where the interaction with the channel boundaries is neglected.
Surprisingly, CSM  of soft particles can be driven also by gravitational effects,  whereby the migration direction depends 
on relative directions between the flow  and the gravitational force \cite{Laumann:2017.2}.
Migration in Newtonian fluids was also found for non-symmetric soft particles in time-periodic linear shear flows 
\cite{Laumann:2017.1} and  even shaken liquids 
when  particle inertia is considered \cite{Kanso:2016.1}. 

Recent studies of particle CSM use besides Newtonian carrier fluids also
visco-elastic fluids. They
also break the fore-aft symmetry and may cause already  CSM of rigid particles.\cite{Avino:2017.1}
CSM in viscoelastic liquids is often faster than  in Newtonian liquids, which  makes non-Newtonian liquids 
attractive for applications such as in  health 
care or biological and chemical analysis.\cite{DAVINO:2010.1,Dan:2018.1,Giudice:2017.1,Lu:2017.1,Faridi:2017.1,Avino:2017.1} 
Particles in non-Newtonian liquids are sometimes also focused to positions aside of the 
channel center even in  the limit of low Reynolds number flows (see e.g. Ref.~\cite{Giudice:2017.1}).
Since in such non-Newtonian liquids shear thinning, leading to a {\it non-constant viscosity},
comes often simultaneously
into play with a fluid elasticity, the specific contribution
of a non-constant viscosity to particle CSM is not clear.

Here  we study the effects of  viscosity gradients on the flow profiles and on the particle dynamics, 
whereby viscosity gradients may be imposed in a controlled way,  for instance, 
by applying a temperature gradient to fluids    \cite{Vincent:2013.1}.
Our modeling approach is described Sec.~\ref{sec_model}, where also analytical expressions for certain flow
profiles and a generalized Oseen tensor are given. 
In Sec.~\ref{sec_explanation}  and Sec.~\ref{sec_shearflow} we show by 
symmetry arguments and numerical simulations,  how  a
 viscosity varying perpendicular to the stream lines of a linear shear   breaks symmetries 
 and induces CSM of capsules already in simple shear flows, in contrast to
 liquids with constant viscosity.
 Two types of viscosity profiles are investigated for plane Poiseuille flows
 in Sec.~\ref{sec_Poiseuilleflow}, where we find also a new scenario for particle stream splitting with 
 interesting applications for particle sorting.

%
%=========================================================================
\section{Models and Methods}\label{sec_model}
%============================================================================================
In Sec.~\ref{models_flow} we consider 
a  constant viscosity gradient perpendicular to the flow lines
in linear shear flows and plane Poiseuille flow. 
We  provide for both cases analytical expressions for the flow profile as solutions of 
the Stokes equation with non-constant viscosity.
In Sec.~\ref{models_particles}
we present a generalized Oseen tensor, which takes
the first correction of the viscosity gradient into account. It is used in the 
Stokesian dynamics simulations of the capsule and is derived without the 
hydrodynamic capsule-wall interactions. 
The wall effects  are taken into account
by the Lattice Boltzmann Method described in 
Sec.~\ref{LBMethod}.

\subsection{Stokes-flows}\label{models_flow}
%============================================================================================
We consider fluids between two boundaries located at $y=0,d$ and 
a spatially varying viscosity
\begin{eqnarray}
\eta({\bf r})=\eta_0+{\bf G}_{\mathrm{\eta}}\cdot {\bf r}
\end{eqnarray}
with a constant gradient vector
\begin{align}
 \nabla \eta ={\bf G}_\eta\,.
\end{align}
We investigate low Reynolds number flows that are determined by the  
Stokes equation
\begin{align}
-\nabla p +\nabla\cdot\left\{\eta [\nabla \bs u+ (\nabla \bs u)^T]\right\}
%+ \bs f
=&0\,\,,
\label{eq_stokes}
\end{align}
with the pressure $p$ and two choices of boundary conditions at $y=0,d$.  

For a {\it classical  shear cell} with one moving boundary the flow field fulfills the boundary conditions:
\begin{align}
\label{ubs}
 {\bf u}(y=0)=0 \quad \mbox{and} \quad{\bf u}(y=d)&=U {\bf e}_x\,.  \quad \mbox{(BC I)}
\end{align}
For a  pressure driven plane Poiseuille flow with a constant  pressure gradient
in $x$-direction, $\nabla p=p_0{\bf e}_x$,
we use the  boundary conditions
\begin{align}
\label{ubp}
 {\bf u}(y=0,d)=0\,.  \hspace{4cm} \mbox{(BC II)}
\end{align}
If not stated otherwise, we consider further on a viscosity gradient in $y$-direction 
\begin{align}
\label{eq_eta}
\eta({\bf r})=\eta_0+G_{\mathrm{\eta},y} y
\end{align}
which may be imposed, for instance, by a temperature gradient perpendicular to the two bounding plates.
For the viscosity gradient, $G_{\eta,y}\not =0$, and the solution of the Stokes equation (\ref{eq_stokes}) 
for the boundary conditions BC~I gives a nonlinear $y$-dependence
of the velocity in $x$-direction
\begin{eqnarray}\label{eq_nonlinear_flow}
{\bf u}_0(y) = U \frac{\ln[y\,G_{\mathrm{\eta,y}}/\eta_0+1]}{\ln[d\,G_{\mathrm{\eta,y}}/\eta_0 +1]} {\bf e}_{\mathrm{x}}\,.
\end{eqnarray}
It reduces in the limit $G_{\eta,y}=0$ to the well known  linear shear profile
\begin{eqnarray}\label{eq_linear_flow}
{\bf u}_0(y) = U \frac{y}{d} {\bf e}_{\mathrm{x}}\,.
\end{eqnarray}
For a  pressure driven flow between two flat boundaries with the boundary conditions 
in Eq. (\ref{ubp}) the $y$-dependence of the flow ${\bf u}_0(y)$ parallel to the $x$-axis is given by
\begin{align}\label{eq_P_flow_grad_y}
{\bf u}_0(y) &= U \frac{Cy-d\ln\left(\frac{G_{\mathrm{\eta,y}}y}{\eta_0}+1\right)}{d\left[1+\ln\left(\frac{C\eta_0}{d\,G_{\mathrm{\eta,y}}}\right)\right]-\frac{C\eta_0}{G_{\mathrm{\eta,y}}}} {\bf e}_{\mathrm{x}}\,, 
\nonumber \\
\mbox{with} \quad C&=\ln\left(\frac{d\,G_{\mathrm{\eta,y}}}{\eta_0}+1\right)\,.
\end{align}
This gives in the limit $G_{\eta,y}=0$ the well known
parabolic profile  ${\bf u}_0(y)= 4 Uy d(d-y) {\bf e}_x$\,.

%============================================================================================
\subsection{Stokesian particle dynamics}\label{models_particles}
%============================================================================================

The surface of the capsule is discretized with $N$ beads at the positions ${\bf r}_i ~(i=1, \ldots, N$).
Their Stokesian  dynamics is described by\cite{Dhont:96}
\begin{eqnarray}\label{beaddynamics}
{\bf\dot r}_i = {\bf u}_0({\bf r}_i) + \sum \limits_{j=1}^N {\bf H}_{ij} \cdot {\bf F}_j~.
\end{eqnarray}
 The capsule center is given by ${\bf r}_{\mathrm{c}} = \sum_{i=1}^{N} {\bf r}_i/N$.
The force on bead $j$ is calculated via ${\bf F}_j = -\nabla_j V({\bf r})$
with $V({\bf r})$ denoting the total potential (given in the following) and ${\bf{ H}}_{ij}$ means the mobility matrix.
The mobility matrix is given by
\begin{eqnarray}\label{mobility}
{\bf H}_{ij}=
\begin{cases}
      \frac{1}{6\pi\eta_i a}\,{\bf 1} & \mbox{ if } i = j \,,\\
       {\bf O}( {\bf r}_i, {\bf r}_j) & \mbox{otherwise} \,.
\end{cases}
\end{eqnarray}
with the  Oseen tensor ${\bf O}$, the Greens function to the Stokes equation (\ref{eq_stokes}).
For a spatially varying viscosity, i. e.  ${\bf G}_\eta\not =0$,
we take  the leading correction with respect to the small quantity
$\frac{({\bf r}_i-{\bf r}_j)\cdot {\bf G}_{\mathrm{\eta}}}{\eta_j}$ 
  into account
\begin{eqnarray}\label{oseen}
{\bf O}({\bf r}_i, {\bf r}_j  )=\frac{1}{8\pi\eta_j R_{i,j}}\left[\bigg(1-\frac{{\bf R}_{i,j}\cdot {\bf G}_{\mathrm{\eta}}}{2\eta_j}\bigg)\right. \nonumber\\ \bigg({\bf 1}+\hat{\bs R}_{i,j}\hat{\bs R}_{i,j}\bigg)
\left.+\frac{1}{2\eta_j}\bigg( {\bf R}_{i,j}{\bf G}_{\mathrm{\eta}}- {\bf G}_{\mathrm{\eta}} {\bf R}_{i,j}\bigg)\right]\,.\nonumber\\
\label{ossen_grad_eta}
\end{eqnarray}
Herein we  use
$\eta_j=\eta({\bf r}_j)$,  ${\bf R}_{i,j}={\bf r}_i-{\bf r}_j$, $r_{i,j}=|{\bf R}_{i,j}|$ 
and $\hat{\bf R}_{i,j}=\frac{\bs R_{i,j}}{r_{i,j}}$. A small value of 
$\frac{({\bf r}_i-{\bf r}_j)\cdot {\bf G}_{\mathrm{\eta}}}{\eta_j}$ 
 means that the spatial deviation of the viscosity on the size of the capsule is small compared to the 
 local viscosity at the position of the capsule. It can be estimated by  the dimensionless number
\begin{equation}
\widetilde{\mbox{\bf G}}_{\mathrm{\eta}}=\frac{{\bf G}_{\mathrm{\eta}} 2R_{\mathrm{c}}}{ \eta_{\mathrm{c}}}
\label{eq_dimless_G}
\end{equation}
with the  viscosity at the center of the capsule $\eta_{\mathrm{c}}=\eta({\bf r}_{\mathrm{c}})$
and the capsule's radius $R_{\mathrm{c}}$. In this form of the Oseen tensor the interaction with the walls
is neglected. The derivation of the expression in  Eq.~(\ref{ossen_grad_eta}) is given in SI.

To calculate the forces and the velocity of the capsule on its surface, which is spherical in its equilibrium shape, 
it must be discretized (see Fig. \ref{fig_Sketch_tra}). We begin with a regular icosahedron, 
which has 12 nodes, and refine the surface iteratively \cite{Krueger:2011.1}: 
We add new nodes at the middle of each edge and shift them to the surface of the sphere, 
until we obtain a good resolution. With this discretization we can calculate 
the forces at the surface whereby we use an elastic force, 
a bending force and a penalty force that ensures volume conservation. 
The elastic force is modeled by the neo-Hookean law that describes a 
thin plate with a constant surface shear elastic modulus $G_s$ 
with a potential $V_{NH}$ (for details see Refs.~\cite{ramanujan:1998.1,BarthesBiesel:2016.1}). 
The bending force follows from the potential $V_{b}$ \cite{Gompper:1996.1} with 
$
V_{b} = \frac{\kappa}{ 2} \sum \limits_{i,j} \left( 1 - \cos \beta_{i,j} \right)
$
whereby $\kappa$ describes the bending stiffness and $\beta_{i,j}$ denotes the angles of the normal vectors between two neighboring triangles.
Furthermore we use a penalty force that ensures that the volume is approximately conserved during the simulations \cite{Krueger:2013.1}. 
Its potential is given by 
$
V_{v} = \frac{k_v}{{\cal V}_0} ({\cal V}(t) - {\cal V}_0)^2
$
with the instantaneous volume ${\cal V}(t)$, the reference volume ${\cal V}_0$ and the rigidity $k_v$.
It is useful to measure the capsule's stiffness with a dimensionless number, the capillary number
\begin{equation}
\mbox{Ca}=\frac{\eta_0 R_{\mathrm{c}}}{G_s}\dot\gamma
\label{eq_def_Ca}
\end{equation}
with (mean) shear rate 
\begin{equation}
 \dot \gamma =\frac{U}{d}\,.
\end{equation}

If not stated otherwise we use the following parameters for the Stokesian dynamics.
Parameters of the flow: $d$=50, $U$=0.5, $\eta_0$=3, ${\bf G}_{\mathrm{\eta}}=0.03\,\hat{\bf e}_{\mathrm{y}}$.
Parameters of the capsule:
initial position $x_0$=0, $y_0$=d/2, $z_0$=0, forces: $k_v$=3.0, $\kappa=0.2$, $G_s=0.2$ (linear shear flow) and $G_s=0.4$ (Poiseuille flow),
mean bead distance $b=1.0$, Radius $R_{\mathrm{c}}=6.6$,
bead radius  $a=0.2$, 
time step $\Delta t=0.05$. 
This leads to $\mbox{Ca}\approx1$, $|\widetilde{\mbox{\bf G}}_{\mathrm{\eta}}|=0.18$ 
%if ${\bf G}_{\mathrm{\eta}}\parallel-\hat{\bf e}_{\mathrm{y}}$ and $|\widetilde{\mbox{\bf G}}_{\mathrm{\eta}}|=0.13$ if if ${\bf G}_{\mathrm{\eta}}\perp\hat{\bf e}_{\mathrm{y}}$ 
at the initial position.

A conversion of the parameters to SI units is obtained by multiplying them with: 
\begin{equation}
u_{\mathrm{m}}=37.88\,\mu\mbox{m,}\ \ u_{\mathrm{s}}=1.89\,\mbox{ms,}\ \ u_{\mathrm{kg}}=2.39\cdot 10^{-11}\,\mbox{kg}\,.
\label{eq_umrechnung_SI}
\end{equation} 
The radius of the capsule is $R_{\mathrm{c}}\approx250\,\mu\mbox{m,}$ and 
the plate distance is $d\approx2\,\mbox{mm}$. 
The viscosity of the fluid at the boundaries (if ${\bf G}_{\mathrm{\eta}}\parallel-\hat{\bf e}_{\mathrm{y}}$) corresponds 
to water at $20^\circ C$ with $\eta(y=0)=1\,\mbox{mPas}$ and $60^\circ C$
with $\eta(y=d)=0.5\,\mbox{mPas}$ \cite{Kestin:2009.1,Gupta:2014.1}. 
This is a temperature gradient comparable to one discussed in Ref.~\cite{Vincent:2013.1}. The maximal velocity is $U=1\,\mbox{cm / s}$.

%============================================================================================
\subsection{The lattice-Boltzmann method \label{LBMethod}}
%============================================================================================
To investigate the particle dynamics without the constraint 
of a small viscosity gradient as for the Stokesian dynamics and in order to take also the
effects of the boundaries on particle dynamics into account we use the lattice-Boltzmann method (LBM). 

\begin{figure}[htb]
	\begin{center}
\includegraphics[width=0.95\columnwidth]{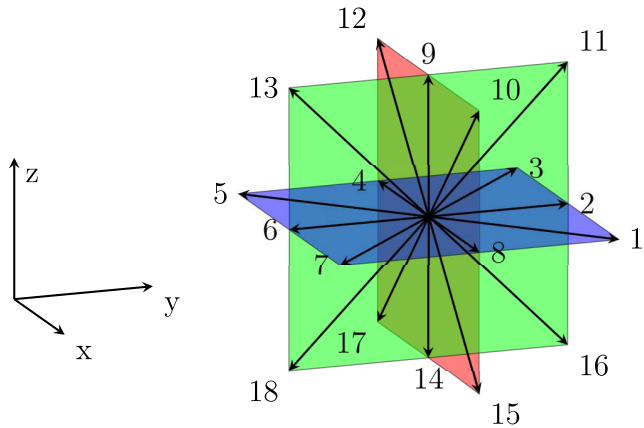}
\end{center}
	\caption{Sketch of the discretized velocity directions of the D3Q19 model for lattice-Boltzmann simulations.}
	\label{fig:D3Q19}
\end{figure}

We use a LBM with 19 discrete velocity directions (D3Q19), cf. fig. \ref{fig:D3Q19} \cite{KruegerT:2016}, 
with the Bhatnagar-Gross-Krook (BGK) collision operator \cite{Bhatnagar:1954.1,Aidun:2010.1}. 
The equation of the probability distribution $f_i(\mbs r, t)$ in velocity-direction $i$ at  position $\mbs r$ is then given by
\begin{eqnarray}
f_i ( \mbs r + c_i \Delta t, t +\Delta t) &=& f_i (\mbs r,t)  -\frac{\Delta t}{\tau} \left( f_i (\mbs r,t) 
- f^{\mathrm{e}}_i (\mbs r,t)\right)\nonumber\\ 
&+& \Delta t F_i\,,
\end{eqnarray}
whereby $\tau$ is a typical relaxation time related to the viscosity of the fluid and
$F_i$ contains the external forces\cite{Guo:2002.1}. 
 $f_i^{\mathrm{e}} (\mbs r,t)$ is the equilibrium distribution:
\begin{eqnarray}
	f^{\mathrm{e}}_i (\mbs r,t) &\approx& \rho w_i \left[ 1 + \frac{\left(\mbs c_i \cdot \mbs u \right)}{c_{\mathrm{s}}^2} + \frac{\left(\mbs c_i \cdot \mbs u\right)^2}{2c_{\mathrm{s}}^4} - \frac{ u^2}{2c_{\mathrm{s}}^2} \right]\nonumber\\
	 &+& \mathscr{O}(u^3)
\end{eqnarray}
with the unit vectors $c_i$ along the discrete directions for the $i$-th velocity. 
Furthermore we use the equilibrium fluid density $\rho_0$, the speed of sound 
in the LBM-system, $c_{\mathrm{s}} = \frac{1}{\sqrt 3}$, and the weighting factors $w_i$ \cite{Aidun:2010.1}.

The probability distribution function allows to calculate the density and the velocity of the fluids via
\begin{eqnarray}
	\rho(\mbs r, t) &=& \sum \limits_i f_i(\mbs r, t) \,,\\
	\rho(\mbs r,t) \mbs u(\mbs r,t) &=& \sum \limits_i \mbs c_i f_i( \mbs r,t) +\frac{1}{2}\Delta t \mbs F(\mbs r)\,,
\end{eqnarray}
whereby $\mbs F(\mbs r)$ is the external force density\cite{Guo:2002.1}. The viscosity of the fluids is given by
\begin{eqnarray}
	\nu(\mbs r) = c_{\mathrm{s}}^2 \left( \tau(\mbs r) - \frac{1}{2} \right)\Delta t\,.
\end{eqnarray}
We use a spatial dependent $\tau(\mbs r)$ to simulate the viscosity gradient given  by Eq. (\ref{eq_eta}).

The external forces are coupled to the flow via the immersed boundary method \cite{Peskin:2002.1}.
Thereby one has to consider that the nodes on the membrane of the capsule do not lie on the discrete grid points of the fluid. 
 The force acting on a node of the capsule's surface is distributed to neighbouring fluid nodes with the function 
$\phi(\Delta \mbs r) = \tilde \phi(\Delta x) \tilde\phi ( \Delta y) \tilde\phi(\Delta z)$ and
\begin{eqnarray}
	\tilde \phi(R) = \left\{ \begin{array}{c l}  \frac{1}{4} \left( 1 + \cos( \frac{\pi R}{2} ) \right) & \text{ if } \left|R\right| \leq 2 \\ 0 & \text{ else } \end{array}\right. \,.
\end{eqnarray}
%This function is used to distribute the force acting on the capsule's surface to the neighboring fluid grid nodes. 
It is also utilized to calculate the velocity at the nodes of the capsule's surface with the velocity of the neighboring fluid nodes.
We use periodic boundary conditions in $x$ and $z$-direction and a standard 
bounce back scheme at the walls to drive the flow \cite{Aidun:2010.1}.

We use the following parameters for the linear shear flow:
Parameters of the flow: 
density $\rho_0$=1.0, 
viscosity $\eta_0=1/6$ at $x=0,~y=0,~z=0$, 
viscosity gradient ${\bf G}_{\mathrm{\eta}}=0.002\,\hat{\bf e}_{\mathrm{y}}$,
velocity of the upper boundary or maximum velocity $U$=0.005, 
number of nodes in $x$-direction $N_x$=400, 
number of nodes in $z$-direction $N_z$=100,
wall distance is $d=100$.
Parameters of the capsule:
initial position $x_0$=0, $y_0$=51.5 and $z_0$=49.5, 
coefficient of the volume preserving force  $k_v$=0.01, 
bending potential $\kappa=10^{-4}$, 
neo-Hookean coefficient $G_{\mathrm{s}}=10^{-4}$ node distance $b=1.0$, which leads to a radius $R=6.6$, 
number of nodes $N=642$ or $R=13.2$ with $N=2562$.
The time step is $\Delta t=1.0$. 
For comparison between LBM and Stokesian dynamics simulations besides the same parameters  a bead radius a=$0.2$ is used. 

We use the following parameters for the Poiseuille flow.
Parameters of the flow: 
density $\rho_0$=1.0, viscosity $\eta_0=1/6$ 
 at $x=0, y=0, z=0$,
viscosity gradient $|{\bf G}_{\mathrm{\eta}}|=0.009$,
velocity of the upper boundary or maximum velocity $U$=0.15, 
number of nodes in $x$-direction $N_x$=400, 
number of nodes in $z$-direction $N_z$=100,
wall distance is $d=300$. Parameters of the capsule:
coefficient of the volume preserving force  $k_v$=0.01, 
node distance $b=1.0$ which leads to a Radius $R=3.3$, number of nodes $N=162$ or $R=13.2$ with $N=2562$,
time step $\Delta t=1.0$,
bending potential $\kappa=0.17$ (large $R$),
bending potential $\kappa=0.02$ (small $R$),
neo-Hookean coefficient $G_{\mathrm{s}}=10^{-3}$ (large $R$),
neo-Hookean coefficient $G_{\mathrm{s}}=5\times10^{-4}$ (small $R$).

%============================================================================================
\section{Explanation of $\nabla \eta$-induced CSM}\label{sec_explanation}
%============================================================================================

We develop at first a qualitative explanation of the CSM induced by a viscosity gradient perpendicular to stream lines.
A viscosity gradient modifies  the flow profiles as indicated in Eq.~(\ref{eq_nonlinear_flow}) and Eq.~(\ref{eq_P_flow_grad_y}), 
i.e. the shear rate across the particle is not constant but is slightly varying. 
This variation of the shear rate is neglected for the qualitative explanation here.
We show that the CSM is directly caused by 
the $\nabla \eta$-induced modifications of the friction forces acting on the particle's surface 
and not indirectly by the varying shear rate (this is also confirmed by simulations, cf. SI). 

\begin{figure}[htb]
\vspace{-2mm} 
\begin{center}
\includegraphics[width=0.95\columnwidth]{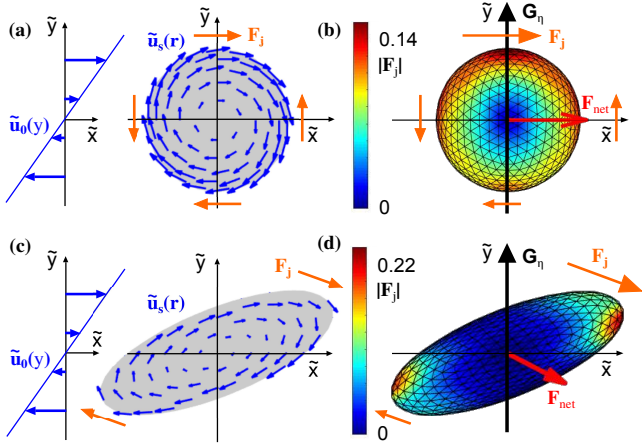}
\end{center}
\vspace{-0.4cm}
\caption{A rigid capsule is rotating due to the shear flow $\tilde{\bf u}_0(y)$ with velocity $\tilde{\bf u}_{\mathrm{s}}({\bf r})$ 
at its surface (comoving frame, without gradient) (a). This leads to friction forces ${\bf F}(\tilde{\bf r})$ (orange arrows), but the sum of these forces is zero because of the symmetry to the $\tilde{x}\tilde{z}$- and $\tilde{y}\tilde{z}$-plane. This motion in presence of a viscosity gradient ${\bf G}_{\mathrm{\eta}}\parallel\hat{\bf e}_{\mathrm{y}}$ (black) leads to higher friction forces on one half than on the other (orange arrows and color of surface) (b). This asymmetry causes a net force ${\bf F}_{\mathrm{net}}$ (red) which is oriented in flow direction, i.e. it causes no CSM.
A soft capsule is deformed and performs a tank-treading motion (c) in a linear shear flow (shown without a gradient). Due to its ellipsoidal shape it is not symmetric to the $\tilde{x}\tilde{z}$- and $\tilde{y}\tilde{z}$-plane, but has a point symmetry that prevents a net force. A gradient (d) breaks the point 
symmetry and leads to a net force with a component perpendicular to the flow. This results in a CSM towards regions with a lower viscosity.}
\label{fig_explanation_x}
\end{figure}

We consider at first a  spherical capsule. 
Without a viscosity gradient, the capsule rotates due to the linear shear
flow (see Eq. (\ref{eq_linear_flow}) and Fig. \ref{fig_explanation_x} (a)) 
and its center ${\bf r}_\mathrm{c}$ follows a flow line. \cite{Dhont:96,Bagchi:2008.1,BarthesBiesel:1980.1}
The velocity at the capsule's surface in the comoving frame 
is given by
$ \tilde{\bf u}_{\mathrm{s}}({\bf \tilde{r}})={\boldsymbol \omega}\times{\bf \tilde{r}}\,
% \label{eq_rot}
$
with  ${\bf \tilde{r}}={\bf r}-{\bf r}_{\mathrm{c}}$, $\boldsymbol \omega=\frac{1}{2}\nabla\times{\bf u}_0$ and the shear flow 
in the comoving frame 
%\begin{equation}
 $\tilde{\bf u}_0(\tilde{y})= U \tilde{y}/d {\bf e}_{\mathrm{x}}\,.$
% \label{eq_u_comoving}
%\end{equation}
The friction forces ${\bf F}({\bf \tilde{r}})$ between the capsule and the fluid can be calculated
by solving Eq. (\ref{beaddynamics}) for the forces ${\bf F}_j ={\bf F}({\bf \tilde{r}}_j)\,$. 
In the following we show how this friction force is affected if this rotation is performed in the presence of a viscosity gradient.

A spherical capsule in a linear shear flow without viscosity gradient has symmetries: 
%It does not change if it is mirrored at the $xz$- or $yz$-plane and its sign is changed subsequently 
The spherical capsule is invariant under a reflection at the $\tilde{x}\tilde{z}$- or $\tilde{y}\tilde{z}$-plane. 
Also the flow's magnitude is equal after these reflections, but the flow changes its sign, 
cf. Fig. \ref{fig_explanation_x} (a). Therefore the velocity and the friction at the surface of the capsule 
have the same symmetries: 
At the mirrored position ${\bf \tilde{r}}'$ of ${\bf \tilde{r}}$ at the $\tilde{x}\tilde{z}$-plane 
we get $F_{\mathrm{x}}({\bf \tilde{r}}')=-F_{\mathrm{x}}({\bf \tilde{r}})$ and $F_{\mathrm{y}}({\bf \tilde{r}}')=F_{\mathrm{y}}({\bf \tilde{r}})$ 
(analogously at the $\tilde{y}\tilde{z}$-plane). The direction of the friction forces from the fluid on the capsule is indicated in Fig. \ref{fig_explanation_x} (a). 
As example at the point of the capsule with the highest $y$-value the friction force points in positive $x$-direction 
and at the point with the lowest $y$-value the force points in negative $x$-direction and has the same magnitude.
This symmetry determines the net force via
\begin{align}
F_{\mathrm{net,x}}&=\oint F_{\mathrm{x}}dA=\int_{y>y_{\mathrm{c}}} F_{\mathrm{x}} dA+\int_{y<y_{\mathrm{c}}} F_{\mathrm{x}} dA\nonumber \\
&= \int_{y>y_{\mathrm{c}}} F_{\mathrm{x}} dA-\int_{y>y_{\mathrm{c}}} F_{\mathrm{x}} dA=0\label{eq_fnet_x_sym}\,,\\
F_{\mathrm{net,y}}&=\oint F_{\mathrm{y}}dA=\int_{x>x_{\mathrm{c}}} F_{\mathrm{y}} dA+\int_{x<x_{\mathrm{c}}} F_{\mathrm{y}} dA\nonumber\\
&=\int_{x>x_{\mathrm{c}}} F_{\mathrm{y}} dA-\int_{x>x_{\mathrm{c}}} F_{\mathrm{y}} dA=0 \label{eq_fnet_y_sym}\,,
\end{align}
whereby $\int_{y>y_{\mathrm{c}}}dA$ denotes an integration over the half sphere on the side of the $xz$-plane with $y>y_{\mathrm{c}}$. 
The symmetries show that the force on one half of the sphere has the opposite direction 
of the  force on the other half. Thus the net forces is zero for a constant viscosity.
Furthermore the system is symmetric to the $\tilde{x}\tilde{y}$-plane which prevents a force in $z$-direction:
\begin{align}
F_{\mathrm{net,z}}&=0\,. \label{eq_fnet_z_sym}
\end{align}
We discuss now the effect of a viscosity gradient oriented perpendicular to the flow direction and 
in the shear plane, i. e. ${\bf G}_{\mathrm{\eta}} \parallel\hat{\bf e}_{\mathrm{y}}$ as shown in Fig. \ref{fig_explanation_x} (b). With the viscosity gradient the friction at the upper half of a rigid spherical capsule ($y>y_\mathrm{c}$), 
which is directed in positive $x$-direction, is higher due to the higher viscosity 
than at the lower half ($y<y_\mathrm{c}$), which is directed in negative $x$-direction. Because the magnitude of the friction is not equal at both halves the symmetry 
used to derive Eq. (\ref{eq_fnet_x_sym}) is broken. Thus a net force is caused by the rotation in presence of the viscosity gradient, even in a linear shear flow.

But the magnitude of the friction still has a symmetry to the $\tilde{y}\tilde{z}$-plane. This can be seen with Fig. \ref{fig_explanation_x} (b) 
by comparing the left part of the capsule ($x<x_\mathrm{c}$) and the right part ($x>x_\mathrm{c}$).
Both halves are symmetric because the viscosity increases in $y$-direction and not in $x$-direction. 
Thus the symmetry used in Eq. (\ref{eq_fnet_x_sym}) is broken, 
but eqs. (\ref{eq_fnet_y_sym}) and (\ref{eq_fnet_z_sym}) are still valid.
Therefore, the net force is oriented in $x$-direction, i.e. ${\bf F}_{\mathrm{net}}$
is parallel to the flow. Thus a rigid sphere
shows migration along the flow direction but no CSM. 
The effects of further possible directions of the viscosity gradient on rigid particles are discussed in SI.
%With this considerations we find that the orientation of the net force is in general given by ${\bf F}_{\mathrm{net}}\parallel \omega\times{\bf G}_{\mathrm{\eta}}$

%If the viscosity gradient is oriented perpendicular to the flow ${\bf G}_{\mathrm{\eta}} \parallel\hat{\bf e}_{\mathrm{y}}$ the situation is similar (see Fig. \ref{fig_explanation_x} (c)): The symmetry used in eq. (\ref{eq_fnet_x_sym}) is broken by the gradient but not eqs. (\ref{eq_fnet_y_sym}) and (\ref{eq_fnet_z_sym}). Thus the resulting force is parallel to the flow. This leads to a migration in the flow's direction, which means no CSM is found.
%If the viscosity gradient is oriented perpendicular to the shear plane ${\bf G}_{\mathrm{\eta}} \parallel\hat{\bf e}_{\mathrm{z}}$ $F_{\mathrm{z}}$ can be locally nonzero. But non of the eqs. (\ref{eq_fnet_x_sym}), (\ref{eq_fnet_y_sym}) and (\ref{eq_fnet_z_sym}) is broken. Thus the symmetry prevents a net force in this case.

The behavior of a deformable capsule is different. Its tank-treading motion and shape obtained by simulations
in a linear shear flow without a viscosity gradient is shown by Fig. \ref{fig_explanation_x} (c). 
The capsule adopts an ellipsoidal shape with its major axis inclined with respect to the flow direction.  The capsule's  center 
follows the  flow direction.
The friction forces are calculated 
in the same way as for a rigid capsule. The main difference to the rigid capsule is the ellipsoidal shape, 
which has no mirror symmetries with respect to the $\tilde{x}\tilde{z}$- and $\tilde{y}\tilde{z}$-plane.
But the deformed shape and the shear flow have both a point symmetry to the capsules center (see Fig. \ref{fig_explanation_x} (c)) and a symmetry to
the $\tilde{x}\tilde{y}$-plane. The friction force has the same symmetry. As example the friction force at two points is shown in Fig. \ref{fig_explanation_x} (c). At the points with the highest $y$-value the friction force from the flow on the ellipsoidal, tank-trading particle points in positive $x$- and negative $y$-direction. At the mirrored points with the lowest $y$-value the force points in negative $x$- and positive $y$-direction.
%At the position ${\bf \tilde{r}}$ and its mirrored position at capsules center ${\bf \tilde{r}}'$ the friction force fulfills $F_{\mathrm{x}}({\bf \tilde{r}}')=-F_{\mathrm{x}}({\bf \tilde{r}})$  and $F_{\mathrm{y}}({\bf \tilde{r}}')=-F_{\mathrm{y}}({\bf \tilde{r}})$. 
Thus eqs. (\ref{eq_fnet_x_sym}), (\ref{eq_fnet_y_sym}) and (\ref{eq_fnet_z_sym}) 
can also be used in case of a deformable capsule, which means ${\bf F}_{\mathrm{net}}=0$ for a constant viscosity.

We discuss now the effect of a viscosity gradient  perpendicular to the flow direction and 
in the shear plane ${\bf G}_{\mathrm{\eta}} \parallel\hat{\bf e}_{\mathrm{y}}$ (other orientations: see SI). The symmetry with respect to the center is broken and eqs. (\ref{eq_fnet_x_sym}) and (\ref{eq_fnet_y_sym}) are not valid in this case. This is shown in Fig. \ref{fig_explanation_x} (d): The force at the point with the highest $y$-value at the high viscosity, which points in positive $x$- and negative $y$-direction, has a higher magnitude than the mirrored force. This leads to a non-zero net force ${\bf F}_{\mathrm{net}}$ which is oriented in positive $x$- and negative $y$-direction.
The system still has a symmetry to the $\tilde{x}\tilde{y}$-plane, so that Eq. (\ref{eq_fnet_z_sym}) is still valid
and the net force has no $z$-component.  The negative $y$-component of the net force leads to a CSM towards the lower viscosity.
Note that this is different to the rigid capsule, whose symmetry to the $\tilde{y}\tilde{z}$-plane prevents a force in $y$-direction. Thus a CSM due to a viscosity gradient is found only if the capsule is soft.

%============================================================================================
\section{CSM in a shear flow \label{sec_shearflow}}
%============================================================================================
Here we confirm by simulations the qualitative reasoning described in the previous section,  
that a finite viscosity gradient, $\nabla \eta$,  causes a CSM of  deformable particles already 
in simple shear flows. 
We use a  generalized Oseen tensor given by Eq.~(\ref{ossen_grad_eta}),
which takes the leading order effects of $\nabla \eta$ into account,  and determine
in Stokesian-dynamics simulations the capsule's CSM velocity as function of  parameters. 
By LBM simulations of a capsule we evaluate the validity range of these approximate
results and we show that capsules in shear flows with $\nabla \eta\not =0$  
are focused to an attractor streamline.

%----------------------------------------------------------------------------------------------
\subsection{Numerical results on $\nabla \eta$-induced bulk migration}
%---------------------------------------------------------------------------------------------
\begin{figure}[htb]
\vspace{-2mm} 
\begin{center}
\includegraphics[width=0.98\columnwidth]{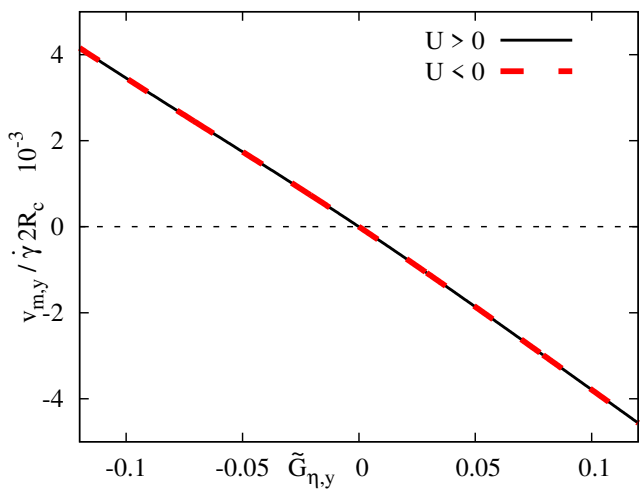}
\end{center}
\vspace{-0.4cm}
\caption{The migration velocity $v_{m,y}$ in units of $2\dot \gamma R_\mathrm{c}$ as function of the dimensionless 
viscosity gradient $\tilde G_{\eta,y}$. 
The CSM   
is directed towards the lower viscosity as sketched in Fig.~\ref{fig_explanation_x}
and it is independent of the sign of $U$, i. e. independent  of the flow direction. 
} 
\label{vm_G_eta}
\end{figure}
In Stokesian dynamics simulations we use the nonlinear shear flow profile given by Eq. (\ref{eq_nonlinear_flow})
and the generalized Oseen tensor given in Eq.~(\ref{ossen_grad_eta}).
We simulate trajectories $y_c(t)$ of the capsule
and determine by linear fits to the slope of capsule trajectories $y_{\mathrm{c}}(t)$ 
the cross-stream migration velocity 
$v_{m,y}$. The resulting $v_{m,y}$ of capsules  
is  shown in Fig.~\ref{vm_G_eta}
as function of the dimensionless viscosity 
gradient $\widetilde{\mbox{G}}_{\mathrm{\eta,y}}$.
The CSM velocity  in Fig.~\ref{vm_G_eta} decreases nearly linearly with $|\widetilde{\mbox{G}}_{\mathrm{\eta,y}}|$ and is oriented as 
explained in the previous section, cf. Fig.~\ref{fig_explanation_x}. In addition,  
the migration is directed towards the lower viscosity and does not depend on the flow's direction, i.e. it is independent on the sign of $U$.
%(Further  orientations of the gradient are discussed in the SI.)
The nonlinear $y$-dependence of the flow velocity in Eq.~(\ref{eq_nonlinear_flow})
and therefore the spatially varying velocity gradient
causes in Fig.~\ref{vm_G_eta} only 
a slight deviation  of $v_{m_y}$ from  a linear dependence on $\tilde{\mbox{\bf G}}_{\mathrm{\eta}}$ at small 
values of  $|\widetilde{\mbox{G}}_{\mathrm{\eta,y}}|$ (a more detailed comparison of the CSM in both flow profiles is given in the SI).
This justifies the assumption  of a constant velocity gradient across the capsule used in the previous section \ref{sec_explanation}.

\begin{figure}[h!]
\vspace{-2mm} 
\begin{center}
\includegraphics[width=0.98\columnwidth]{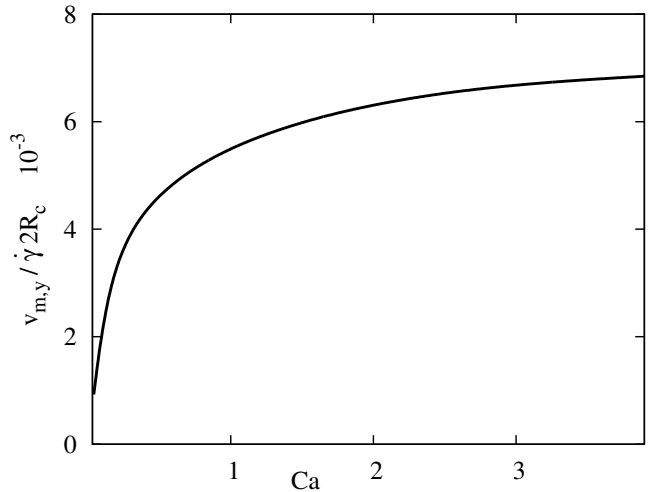}
\end{center}
\vspace{-0.4cm}
\caption{The migration velocity $v_{m,y}$ is given in units of
$2\dot \gamma R_\mathrm{c}$ and as function of the capillary number Ca
given by  Eq. (\ref{eq_def_Ca}).
The migration vanishes at a high capsule-stiffness Ca$\,\ll1$
and increases with Ca. 
} 
\label{vm_G}
\end{figure}
Fig. \ref{vm_G} shows the dependence 
of the migration velocity on the stiffness of the capsule: 
The CSM decreases with increasing stiffness and 
vanishes at small values of the capillary number  Ca$\,\ll1$ given in Eq.~(\ref{eq_def_Ca}). 
This underlines the importance of the deformability of particles for their cross streamline migration (see also  Fig.~\ref{fig_explanation_x}).

%text
%\begin{figure}[htb]
%\vspace{-2mm} 
%\begin{center}
%\includegraphics[width=0.98\columnwidth]{./fig/vm_G_eta_{\mathrm{x}}_vgl.eps}
%\end{center}
%\vspace{-0.4cm}
%\caption{blabla
%} 
%\label{vm_G_eta_{\mathrm{x}}_vgl}
%\end{figure}
\begin{figure}[t]
\vspace{-2mm} 
\begin{center}
\includegraphics[width=0.98\columnwidth]{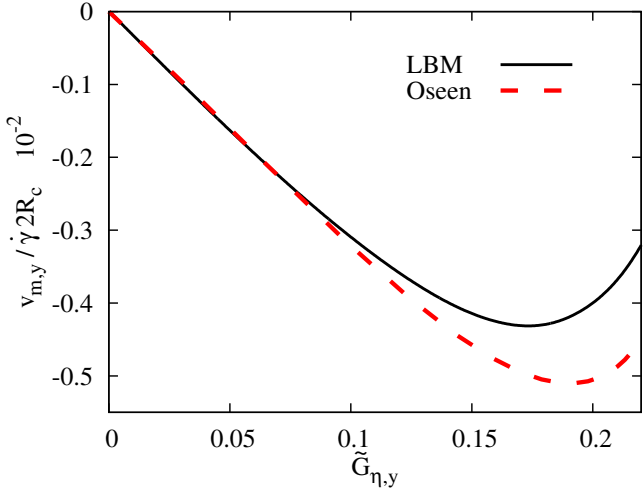}
\end{center}
\vspace{-0.4cm}
\caption{
The CSM velocity $v_{m,y}$ is determined 
by Stokesian dynamics simulations (dashed) and by Lattice-Boltzmann simulations (solid). 
The expansion up to leading order of the
viscosity gradient $\tilde G_\eta$, as used in Stokesian dynamics simulations,
leads for $v_{m,y}$ to an error less than 
$10\%$ if $|\widetilde{\mbox{\bf G}}_{\mathrm{\eta}}|\lessapprox0.15$ compared to the LBM.
} 
\label{fig_vgl_LBM}
\end{figure}

The generalized Oseen tensor in Eq.~(\ref{ossen_grad_eta}) 
takes into account the first order correction with respect to 
the viscosity gradient, i.e.  it is valid  
for small values of $|\widetilde{\mbox{\bf G}}_{\mathrm{\eta}}|$. 
To estimate the validity range of 
%$|\widetilde{\mbox{\bf G}}_{\mathrm{\eta}}|$ 
 this approximation 
we compare the CSM velocity $v_{m,y}$ as obtained by Stokesian dynamics simulations using the
generalized  Oseen tensor in Eq.~(\ref{ossen_grad_eta}) with that  obtained  by   Lattice-Boltzmann simulations of capsules.
In order to keep in LBM simulations the interaction of the capsule with the boundary small, 
we positioned it in the middle of the flow cell between the two boundaries. In addition
we have chosen a small ratio between the capsule's diameter and the wall distance $\frac{2 R_{\mathrm{c}}}{d}\approx0.13$. 
Furthermore a sufficiently small Reynolds number Re=$\frac{\rho U R_{\mathrm{c}}}{\eta_0}\approx0.2$  was chosen in LBM simulations 
to match the low Reynolds number regime of the Stokesian dynamics simulations.
The flow is simulated  for the boundary condition given by Eq. (\ref{ubs}) and  the viscosity  gradient points
into the direction perpendicular to the boundaries.

The migration velocities resulting from both simulations  are shown in Fig.~\ref{fig_vgl_LBM}. 
The simulation results for  the capsule with the generalized Oseen tensor and those obtained 
via the LBM agree well in the range  of small values of $|\widetilde{\mbox{\bf G}}_{\mathrm{\eta}}|$ 
and the deviation increases with $|\widetilde{\mbox{\bf G}}_{\mathrm{\eta}}|$.
For example at $|\widetilde{\mbox{\bf G}}_{\mathrm{\eta}}|\lessapprox0.15$ 
the relative error is below $10\%$ and at $|\widetilde{\mbox{\bf G}}_{\mathrm{\eta}}|\lessapprox0.18$ 
the error is below $20\%$.

%----------------------------------------------------------------------------------------------
\subsection{Particle focusing to an attractor streamline\label{Linshearfocus}}
%----------------------------------------------------------------------------------------------
\begin{figure}[h!]
\vspace{-2mm} 
\begin{center}
\includegraphics[width=0.98\columnwidth]{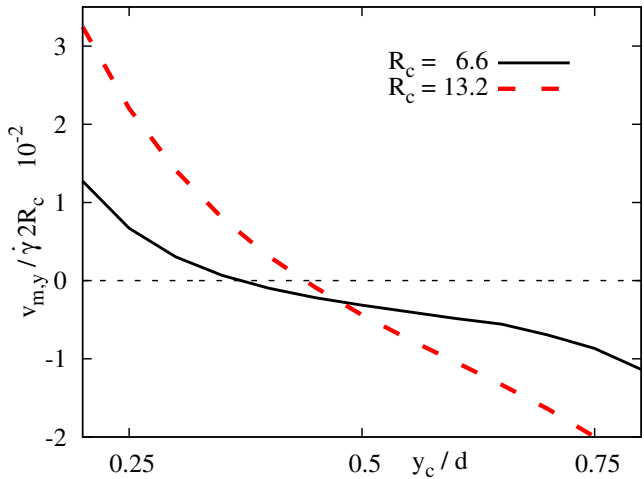}
\end{center}
\vspace{-0.4cm}
\caption{The migration velocity of a soft capsule in a flow with viscosity gradient and  the boundary conditions in Eq.~(\ref{ubs})
is determined by the LBM as function of the initial position $y_c$ for two different particle radii $R_c$. 
Far away from the walls the particle migrates due to the viscosity gradient towards the lower viscosity, 
i.e. towards the plate at $y=0$. Close to the walls the repulsive wall interaction dominates,
which leads to a migration away from the walls. Hence there is a stable position, i.e. an attractor off center, 
that depends on the particle's size. 
It is located at $y\approx0.37d$ with $2R_\mathrm{c}/d=0.13$ and at $y\approx0.41d$ with $2R_\mathrm{c}/d=0.26$.
} 
\label{fig_wall_ineraction}
\end{figure}
The LBM includes the hydrodynamic interaction of the capsule with the walls, which causes a so-called  lift force
that repels the capsule from walls and
that depends on the capsule-wall distance
\cite{Misbah:1999.1,Seifert:1999.1,Viallat:2002.1}.
The interplay with the  lift force causes  a  $y$-dependent migration velocity as shown for two capsules with two different radii
in Fig.~\ref{fig_wall_ineraction}.
The dimensionless gradient ranges in this case 
from $G_{y}=0.16$ (at $y=0$) to $\tilde G_y=0.07$ ($y=d$) with $R_c=6.6$ and with $R_c=13.6$ from $\tilde G_y=0.14$ to $\tilde G_y=0.32$.
%for the capsule radius
%$R_c=6.6$ with $\tilde G_{y}=0.16$ and at $y=0$ and with $\tilde G_y=0.07$ and at $y=d$ 
%and for radius
%$R_c=13.6$ with $\tilde G_y=0.32$ and at $y=0$ and with $\tilde G_y=0.14$ and at $y=d$.
%
The CSM caused by $\nabla \eta$ and the wall repulsion  balance each other in the range of 
the lower viscosity and at this value of $y$ the migration velocity $v_{m,y}$ vanishes.
The location of this attractor position depends on  the capsule size.
We find the attractor at $y\approx0.37d$ for $2R_\mathrm{c}/d=0.13$ and  $y\approx0.41d$ for $2R_\mathrm{c}/d=0.26$.

%----------------------------------------------------------------------------------------------
\section{Simulation of CSM in Poiseuille flow \label{sec_Poiseuilleflow}}
%----------------------------------------------------------------------------------------------

Capsules and red blood cells migrate in a Poiseuille flow, driven by 
the spatial varying shear gradient across a soft particle, usually 
to the center of the flow channel. \cite{BarthesBiesel:1980.1,Kaoui:2008.1,Misbah:2008.1,Bagchi:2008.1}
If one has a viscosity gradient perpendicular to the boundaries
across a plane Poiseuille flow, the $\nabla \eta$ induced migration
has in the whole cell the same direction i. e. the $\nabla \eta$ induced 
migration either supports or acts against  the common center directed migration.
This interplay is investigated by Stokesian dynamics simulation 
in unbounded (bulk) Poiseuille flows  and by LBM simulations, where boundary effects are included.

If a constant $\nabla\eta$ is used, e.g. induced by  a temperature gradient across the flow, 
then the maximal velocity  of a Poiseuille flow is shifted towards the lower viscosity. 
We study here the migration in such a flow profile. However, also with shear thinning fluids a viscosity gradient can be generated.
It is well known from shear thinning fluids, that the viscosity has its maximum in the center of a Poiseuille flow and 
decreases  towards the walls.  Particle migration is recently studied also in Non-Newtonian fluids,
whereby in these works besides shear-thinning effects also elastic effects are considered
to be important. In order to contribute to the understanding of CSM of soft particles in 
 shear thinning fluids, we mimic also shear thinning fluids by studying 
the effects of a viscosity on the migration behavior of a capsule, where the viscosity 
has its maximum in the channel center and decays linearly to the boundaries.

%%%%%%%%%%%%%%%%%%%%%%%%%%%%%%%%%%%%%%%%%%%%%%%%%%%
\subsection{Migration in {\it unbounded} Poiseuille flow induced by $\nabla \eta=const.$}
%%%%%%%%%%%%%%%%%%%%%%%%%%%%%%%%%%%%%%%%%%%%%%%%%%%
Here we consider as in the previous section a  capsule in a fluid with constant viscosity gradient along the $y$-axis 
( e.g. generated by a temperature gradient) given by
Eq.~(\ref{eq_eta}) but now with a Poiseuille flow profile given by   Eq.~(\ref{eq_P_flow_grad_y}). We simulate
 the capsule's Stokesian dynamics  by using the  generalized Oseen tensor given in Eq.~(\ref{ossen_grad_eta}).
With the flow profile in Eq.~(\ref{eq_P_flow_grad_y}) the  simulations focus on the behavior of the capsule in the 
bulk of  a Poiseuille flow, where the hydrodynamic interactions between the capsule and the
wall are negligible. This allows a direct comparison between 
the well known  bulk CSM  in  Poiseuille flow (see e.g. Refs.~\cite{Kaoui:2008.1,Bagchi:2008.1})
and the $\nabla \eta$ induced CSM.

\begin{figure}[htb]
\vspace{-2mm} 
\begin{center}
\includegraphics[width=0.98\columnwidth]{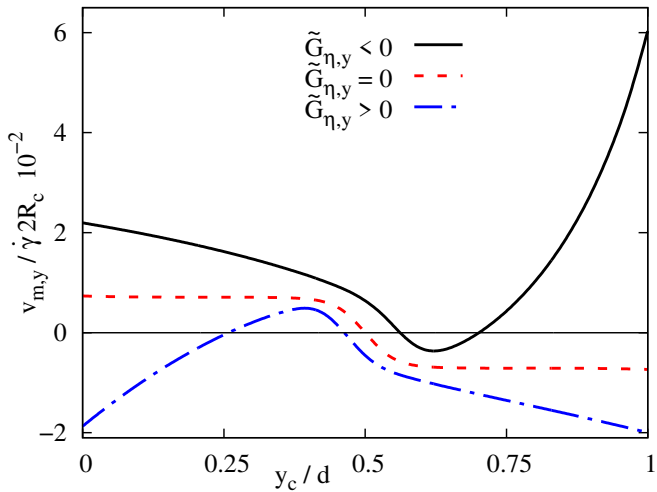}
\end{center}
\vspace{-0.4cm}
\caption{The migration velocity $v_{m,y}/(2\dot \gamma R_\mathrm{c})$  of a capsule in the 
distorted Poiseuille flow profile given by Eq.~(\ref{eq_P_flow_grad_y}) is shown. It is
obtained by Stokesian dynamics simulation  as
function of the capsule position $y/d$.  
The CSM is calculated for two viscosity gradients, one pointing to negative and the other into the positive $y$-direction 
as well as with a vanishing viscosity gradient. 
Without a gradient the capsule migrates towards the center as expected (see e.g. Refs.~\cite{Kaoui:2008.1,Bagchi:2008.1}).
In the case of a viscosity gradient the capsule migrates again towards the lower viscosity, 
besides a small region close to the center, where
the shear rate of the flow profile in Eq.~(\ref{eq_P_flow_grad_y}) vanishes.
} 
\label{vm_G_eta_pflow}
\end{figure}

Figure \ref{vm_G_eta_pflow} shows the migration velocity of the capsule
as function of its $y$ position with and without a viscosity gradient. 
For $\widetilde{\mbox{ G}}_{\mathrm{\eta},y}=0$
the capsule migrates to the center and the related CSM velocity $v_{m,y}$ is indicated by the dashed line.

With a gradient in negative $y$-direction, i.e. $\widetilde{\mbox{G}}_{\mathrm{\eta,y}}<0$, 
the viscosity ranges from $\eta(y=0)=3$ to $\eta(y=d)=1.5$. 
The CSM velocity for this case is given by the solid line in Fig.~\ref{vm_G_eta_pflow}.
The $\nabla \eta$ effect  enhances in  a range of smaller $y$ the CSM velocity to the center, 
i.e. in positive $y$-direction. Near the channel center at $y=0.54d$ the flow has its maximal velocity and 
the shear rate of the flow field given by   Eq.~(\ref{eq_P_flow_grad_y})
 vanishes.  At this position  the $\nabla \eta$ induced migration vanishes 
 too and the migration directed to the channel center dominates. Thus capsules with an initial position $y_0\lessapprox0.7\,d$ migrate until they 
reach the attractor near the center. At  initial positions $y_0\gtrapprox 0.7\,d$ the $\nabla \eta$ 
induced outward migration dominates and the capsules migrate away from the center. 
This outward  migration is near $y=d$ approximately up to 8 times faster than the center
oriented one.

With a gradient in positive $y$-direction the viscosity ranges from $\eta(y=0)=3$ to $\eta(y=d)=4.5$. 
The situation is similar and the migration is also directed to the  region of the 
lower viscosity, which is now located at the plate at $y=0$. Therefore we get $v_{m,y}<0$ again 
besides the region close to the center. 
Capsules with an initial position $y_0\gtrapprox 0.25\,d$ migrate to the attractor close 
to the center at $y=0.44\,d$ and capsules with $y_0\lessapprox 0.25\,d$ 
migrate to the wall at $y=0$. 

%Thus the direction of the migration, 
%i.e. to the center or to the wall, can be controlled by the direction of the gradient 
%if the initial position of the capsule is not too close to the center.

%%%%%%%%%%%%%%%%%%%%%%%%%%%%%%%%%%%%%%%%%%%%%%%%%%%%%%%%%%%%%%%%
\subsection{Migration in {\it bounded} Poiseuille flow induced by $\nabla \eta=const.$}
\begin{figure}[htb]
\vspace{-2mm} 
\begin{center}
\includegraphics[width=0.98\columnwidth]{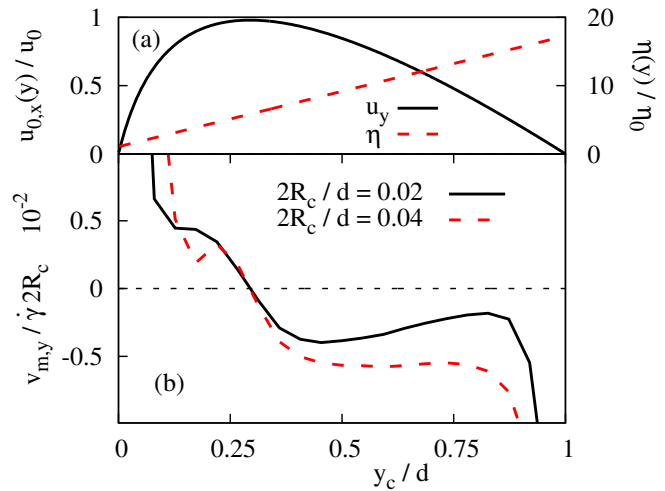}
\end{center}
\vspace{-0.4cm}
\caption{(a) The spatial dependence of the flow velocity $u_{0,x}(y)$  (solid line) and the viscosity for a linear increase 
of  the viscosity between both boundaries (dashed line). The maximal velocity is shifted towards the region with the lower viscosity. 
(b) The spatial dependence of the migration velocity with wall interaction for two different capsule radii $R_c$. 
The attractor with vanishing $v_{m,y}$ is shifted towards the lower viscosity due to the shift of the maximal flow velocity. 
} 
\label{vm_G_eta_pflow_Temp}
\end{figure}

%%%%%%%%%%%%%%%%%%%%%%%%%%%%%%%%%%%%%%%%%%%%%%%%%%%%%%%%%%%%%%
Here we describe results of  LBM simulations of a capsule for  $\nabla \eta=const.$  and 
the flow field boundary conditions given in Eq.~(\ref{ubp}). 
 Figure \ref{vm_G_eta_pflow_Temp} (a) shows the spatial dependence of the flow velocity and the linear increase of the viscosity. 
 This demonstrates that the maximum flow velocity in Poiseuille flow  is shifted towards the lower viscosity.
 
The Fig.~\ref{vm_G_eta_pflow_Temp} (b) shows 
 the migration velocity $v_{m,y}$ as function of $y_\mathrm{c}/d$. 
In contrast to the case without wall interaction (cf. Fig. \ref{vm_G_eta_pflow}) 
only the attractor at the maximal flow velocity and no repeller is found.
The reason is that wall induced repelling lift force is stronger than the $\nabla \eta$ induced migration to large values of $y$,
even at higher values of $\tilde G_y$. 
The shift of the attractor with $v_{m,y}=0$ to smaller values of $y$ than the channel center has its
origin in the shift of the maximal flow velocity to smaller values of $y$.
Hence, in the presence of walls the capsule migrates always to one attractor that is 
shifted away from the flow center by the constant viscosity gradient. 
However, the migration velocity is larger in the presence of the viscosity gradient. 

%%%%%%%%%%%%%%%%%%%%%%%%%%%%%%%%%%%%%%%%%%%%%%%%%%%%%%%%%%%%%%%%%%%%%%
\subsection{Particle attractor splitting induced by $\nabla \eta\neq const.$}
\begin{figure}[htb]
\vspace{-2mm} 
\begin{center}
\includegraphics[width=0.98\columnwidth]{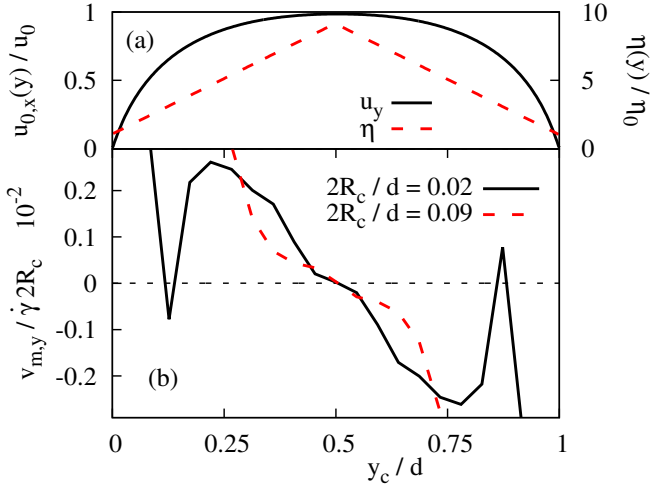}
\end{center}
\vspace{-0.4cm}
\caption{(a) The spatial variations of the shear viscosity (dashed line) and
the flattened velocity profile $u_{0,x}(y)$  (solid line).
(b) The $y$-dependence of migration velocity as obtained by LBM simulations of a capsule for
two different radii. For the smaller particle radius 
a second attractor emerges. 
} 
\label{vm_G_eta_pflow_lin_shear_thinning}
\end{figure}

%%%%%%%%%%%%%%%%%%%%%%%%%%%%%%%%%%%%%%%%%%%%%%%%%%%%%%%%%%%%%%%%%%%%%%
Here we study the capsules dynamics by LBM simulations in   
a viscosity profile that is maximal 
at the channel center and decreases linearly towards the walls, as indicated by the dashed line in Fig.~\ref{vm_G_eta_pflow_lin_shear_thinning} (a). A decay of the shear viscosity 
in Poiseuille flow is known for shear thinning fluids and the viscosity profile in Fig.~\ref{vm_G_eta_pflow_lin_shear_thinning} (a) is a very simple mimicry of
the shear viscosity of shear thinning fluids. 
For this  viscosity profile one obtains in simulations  a Poiseuille flow profile, cf. solid line in Fig.~\ref{vm_G_eta_pflow_lin_shear_thinning} (a),
which is flattened near the channel center similar as  for shear thinning fluids. 

The  migration velocity of  capsules in this viscosity profile 
is shown in Fig.~\ref{vm_G_eta_pflow_lin_shear_thinning} (b)
for two different radii of the capsule. 
The attractor at the channel center is not  shifted by this
viscosity profile, because the shear rate vanishes at the channel center and 
the $\nabla \eta$ induced migration as well. In this region close to the center the center directed migration dominates.
However, for the smaller capsule with $2R_c/d=0.02$ the migration velocity, represented by the solid line
in Fig.~\ref{vm_G_eta_pflow_lin_shear_thinning} (b), 
changes on each side of the 
channel center two times its sign.  At the outer zero of $v_{m,y}$ an additional particle 
attractor  has emerged. It is caused, similar as for the linear shear flow in the previous section \ref{Linshearfocus}, 
by the interplay between outward directed $\nabla \eta$ induced migration, which outweighs 
here the center directed migration, and the wall repulsion. 
This is similar also to the unbounded flow, where the outward directed $\nabla \eta$ induced migration can overcome the center migration 
(cf. Fig. \ref{vm_G_eta_pflow}).
Between two neighboring attractors the vanishing migration velocity $v_{m,y}$ marks a particle repeller.

The emergence of the off-center attractors enables  interesting applications. 
If in a viscosity profile, similar as in Fig.~\ref{vm_G_eta_pflow_lin_shear_thinning} (a), particles of different sizes are injected near one boundary,
 the larger ones migrate to the attractor at the channel center and the smaller ones may stay 
 along the off-center attractor. I. e. at the end of the channel particles of 
 different size or elasticity (cf. Fig. \ref{fig_explanation_x} and Fig. \ref{vm_G})
 are separated.
 This is a interesting new concept in microfluidics  for the separation of different soft particles.

In investigations with viscoelastic fluids  a  particle  migration 
to off-center attractors has been reported before 
\cite{DAVINO:2010.1,Dan:2018.1,Giudice:2017.1,Lu:2017.1,Faridi:2017.1,Avino:2017.1}
and it is not always clear whether this type of  migration is driven more by elastic or viscosity effects.
 Here the mechanisms of an outward directed migration to an off-center attractor, driven by the $\nabla \eta$ effects, 
are rather clear. Therefore, our model with the viscosity profile shown 
Fig.~\ref{vm_G_eta_pflow_lin_shear_thinning} (a),
may help for an improved understanding of CSM in viscoelastic fluids.

%============================================================================================
\section{Discussion and conclusions}
%============================================================================================
We investigated the effects of a spatially varying viscosity on the flow profile in shear and Poiseuille flow
and we described a novel viscosity-gradient driven cross-streamline migration (CSM) of soft capsules, which
 represents deformable particles.  
A viscosity gradient  in microfluidic devices may be induced, for instance, by a temperature gradient \cite{Vincent:2013.1}.

For the Stokesian dynamics simulations of capsules we determined flow profiles that take 
a constant viscosity gradient into account. We also
derived for these simulations a generalized Oseen tensor that includes
the viscosity gradient. These results   may be  also utilized in other approaches such as 
the boundary integral method \cite{Pozrikidis:1992.1} 
or in simulations of microswimmers \cite{Elgeti:2015.1}, polymers \cite{DoiEd} 
and colloids \cite{Dhont:96}. 
%
%We also used Lattice Boltzmann simulations to discuss the wall interaction.
%, e.g. the dynamics of colloids, polymers, vesicles, red blood cells or microswimmers. 
%We derived the Oseen tensor approximately in first order of the viscosity gradient. We compared it therefore with a Lattice Boltzmann method, which solves the full equations without approximations, and we find a good agreement.
%We see that both methods agree well, whereby  the error of the migration velocity less than 6 percent, if the spatial change of the viscosity over the diameter of the capsule is smaller than 15 percent of the viscosity at the particle's center. 

Rigid and soft particles in liquids of constant viscosity do not migrate across the streamlines in linear shear flows \cite{Bagchi:2008.1}. 
We have shown by symmetry arguments  how the interplay between the particle
deformability and the Stokes-friction forces, that vary according to a viscosity gradient
across a particle, leads to
cross-streamline migration of deformable capsules in simple shear flows towards the region of lower viscosity.
This reasoning may also apply to the particle dynamics in non-Newtonian fluid flows, whereby in this case
often elastic effects have to be taken into account as well.
Our prediction on the basis of symmetry arguments are confirmed by Stokesian dynamics simulations. By Lattice Boltzmann
simulations, where  the particle wall interactions are taken into account, we also show that the interplay between
this viscosity-gradient induced migration  and the hydrodynamic wall repulsion causes even in linear shear flows 
a focusing of particles to an attractor streamline  in the low viscosity region as indicated in Fig.~\ref{fig_Sketch_tra}.
The location of the attractor depends on the strength of the viscosity gradient and the particle properties.
This predicted focusing may have interesting applications.

We  investigated  CSM also in  Poiseuille flows  for two different    viscosity gradients. 
A constant viscosity gradient across plane Poiseuille flow may be induced again
by a temperature gradient across a flow cell. CSM in the presence of 
a viscosity gradient is much faster than without a gradient. As example we showed 
that the viscosity gradient,
that  corresponds to water with a temperature difference of $40^\circ$C between  the boundaries at a  distance of 2 $mm$, can 
already enhance the migration velocity by up to a factor 8. Such gradients are reported from experiments \cite{Vincent:2013.1}, 
also higher viscosity gradients can be achieved with e.g.  sucrose in water \cite{Telis:2007.1}. 
Besides the faster migration, also the location of the particle  attractor in Poiseuille flow is affected by the viscosity gradient:
 It is shifted away 
from the center of a Poiseuille flow. 
The major reason for this shift is, that the location of the maximum of flow profile
and  therefore the position of  zero shear rate is shifted towards the 
region of lower viscosity, which also shifts the position of the attractor. Thus the location
of the particle attractor can be controlled by the viscosity gradient
in a Poiseuille flow as well.

Shear thinning fluids   in Poiseuille flows display a variation of the
viscosity gradient with a maximum of the viscosity at the channel center.
We described the viscosity landscape of shear thinning fluids in a simplified manner. 
At the channel center we also have chosen the viscosity maximum and a linear decay towards the channel boundaries.
In order to focus to the effects of viscosity gradients, we have neglected further
possible effects in  complex fluids, such as elastic forces.
The assumed viscosity landscape changed the CSM velocity as function of the distance from channel center
considerably, compared to fluids with  constant viscosity. 
Moreover, the CSM induced by stronger viscosity gradients  dominates
and drives in a larger off-center region of the channel cross section particles towards the boundaries. 
In this range the interplay with particle-wall repulsion may even cause, besides the
particle attractor at the channel center, two further  off-center particle attractors. 
These attractors are found for smaller but not for
larger soft particles. A similar  behavior was also found in experiments with visco-elastic liquids \cite{Avino:2017.1,Giudice:2017.1}.
Here we can identify the appearance of off-center particle attractors in a unique manner
with the viscosity gradient. Our insights may contribute  to a further
understanding of cross-streamline migration in  complex liquids in straight and possibly in
wavy channels \cite{Foerster:2018.1}.

\begin{acknowledgement}
The support by the German French University (DFH/UFA) and discussion with A. F\"ortsch, D. Kienle and W. Schmidt
are gratefully acknowledged.
\end{acknowledgement}

\section*{Author contribution statement}
ML and WZ designed the research;
ML performed the calculations and the simulations;
ML and WZ  interpreted and discussed the results and wrote the paper.

%\footnotesize{
%\bibliography{../../../../bib/poly2,../../../../bib/colloid,../../../../bib/stokes,neu}
\bibliographystyle{prsty}
\bibliography{poly2,colloid,stokes,neu}

%}

\end{document}